\documentclass[a4paper,twoside,reqno]{bjp}

\usepackage{graphicx}
\usepackage{amssymb,amsmath,amscd,amsthm}
\usepackage{times}
\usepackage[bookmarks=false]{hyperref}
\hypersetup{%
    colorlinks=true,        
    linkcolor=blue,          
    citecolor=blue,         
    urlcolor=blue           
    }
\usepackage{geometry}
\allowdisplaybreaks

\usepackage[numbers,square,sort&compress]{natbib}
\usepackage{bebands-sdanca19}

\newcommand{\ifproofpre}[2]{#1}

\begin{document}

\title{\boldmath\textit{Ab initio} rotation in $\isotope[10]{Be}$}

\runningheads{M.~A.~Caprio \textit{et al.}}{\textit{Ab initio} rotation in $\isotope[10]{Be}$}

\begin{start}{%
\author{Mark A. Caprio}{1},
\author{Patrick J. Fasano}{1},
\author{Anna E. McCoy}{2},
\author{Pieter Maris}{3},
\author{James P. Vary}{3}

\address{Department of Physics, University of Notre Dame, Notre Dame, Indiana 46556-5670, USA}{1}
\address{TRIUMF, Vancouver, British Columbia V6T~2A3, Canada}{2}
\address{Department of Physics and Astronomy, Iowa State University, Ames, Iowa
  50011-3160, USA}{3}

}

\begin{Abstract}
%
%
\textit{Ab initio} theory describes nuclei from a fully microscopic formulation,
with no presupposition of collective degrees of freedom, yet signatures of
clustering and rotation nonetheless arise.
%
%
We can therefore look to \textit{ab initio} theory for an understanding of the
nature of these emergent phenomena.
%
%
To probe the nature of rotation in $\isotope[10]{Be}$,
%
%
we examine the predicted rotational spectroscopy from no-core configuration
interaction (NCCI) calculations with the Daejeon16 internucleon interaction,
%
%
and find spectra suggestive of coexisting rotational structures having
qualitatively different intrinsic deformations: one triaxial and the other with
large axial deformation arising primarily from the neutrons.
%
%
%
\end{Abstract}

\begin{KEY}
\textit{Ab initio} nuclear theory, no-core
configuration interaction (NCCI) approach, nuclear rotation, Daejeon16 interaction
\end{KEY}
\end{start}


\newcommand{\fnexptbandmembers}{
  \label{fn:expt-band-members}
The experimental band members for the
  $K^P=0^+_1$ band are $0_1^+$ ($0.00\,\MeV$), $2^+_1$ ($3.37\,\MeV$), and
  $4^+_2$ ($11.76\,\MeV$), which was assigned as $(4^+)$ in
  Ref.~\cite{npa2004:008-010} but confirmed as $4^+$ in
  Ref.~\cite{bohlen2007:10be-pickup}.  Those for the $K^P=0^+_2$ band are
  $0^+_2$ ($6.18\,\MeV$), $2^+_3$ ($7.54\,\MeV$), and $4^+_1$ ($10.15\,\MeV$),
  which was assigned as $3^-$ in Ref.~\cite{npa2004:008-010} but revised to
  $4^+$ in
  Refs.~\cite{freer2006:10be-resonant-molecule,suzuki2013:6he-alpha-10be-cluster}.
  Furthermore, Ref.~\cite{bohlen2007:10be-pickup} identifies a putative $K^P=2^+$ band,
  consisting of $2^+_2$ ($5.96\,\MeV$) and $3^+_1$ ($9.4\,\MeV$), where this
  latter level is newly identified with assignment $(3^+)$ in
  Ref.~\cite{bohlen2007:10be-pickup}.
}

\newcommand{\fnrotational}{
  \label{fn:rotational}
Rotational band members built on an axially symmetric intrinsic state $\tket{\phi_K}$,
with projection $K$ of the angular momentum onto the intrinsic symmetry axis,
have angular momenta $J\geq K$, with
energies $E(J)=E_0+AJ(J+1)$, where the rotational energy constant
$A\equiv\hbar^2/(2\cal{J})$ is inversely related to the moment of inertia
$\cal{J}$, and $E_0=E_K-AK^2$ is related to the energy $E_K$ of the rotational
intrinsic state~\cite{rowe2010:collective-motion}.
}

\newcommand{\fndaejeon}{
  \label{fn:daejeon}
The Daejeon16 internucleon
interaction~\cite{shirokov2016:nn-daejeon16} is obtained starting from the
Entem-Machleidt chiral perturbation theory
interaction~\cite{entem2003:chiral-nn-potl}, which is then softened via a
similarity renormalization group transformation and adjusted via a
phase-shift equivalent transformation to yield an accurate description of light
nuclei with $A\leq16$.
}


\section{Introduction}
\label{sec:intro}

\textit{Ab initio} nuclear theory attempts to provide a fully microscopic and
predictive theory of nuclei, by solving the quantum many-body problem for the
nucleons and their free-space interactions.  Indications of collective phenomena,
including both
clustering~\cite{pieper2004:gfmc-a6-8,neff2004:cluster-fmd,maris2012:mfdn-ccp11,romeroredondo2016:6he-correlations,navratil2016:ncsmc}
and
rotation~\cite{caprio2013:berotor,maris2015:berotor2,caprio2015:berotor-ijmpe},
may be found in the results of \textit{ab initio} calculations.

From their \textit{ab initio} microscopic description, we may seek not only
quantitative predictions, but also qualitative insight into the structure of
these collective degrees of freedom.  \textit{Ab
  initio} calculations provide access to an extensive set of observables,
such as transition strengths, for identifying collective excitations.  The
calculated spectroscopy can go far beyond what is in practice experimentally
accessible for the collective states of these light nuclei, thereby making
collective patterns easier to identify.  Furthermore, the calculated wave functions can
be analyzed as a more direct probe of the nature of their collective
structure~\cite{dytrych2007:sp-ncsm-evidence,dytrych2013:su3ncsm,johnson2015:spin-orbit,caprio20xx:bebands}.

For the odd-mass $\isotope{Be}$ isotopes $\isotope[7,9,11]{Be}$, the nature of
\textit{ab initio} emergent rotation has been probed through such approaches in
Ref.~\cite{caprio20xx:bebands}, based on no-core configuration interaction
(NCCI)~\cite{barrett2013:ncsm} [or no-core shell model (NCSM)] calculations with
the Daejeon16 internucleon interaction~\cite{shirokov2016:nn-daejeon16}.  In the
present contribution, we turn to the even-mass isotope $\isotope[10]{Be}$, which
is experimentally known to have a rich rotational
spectroscopy~\cite{freer2007:cluster-structures} and has been suggested to
exhibit proton-neutron triaxiality~\cite{kanadaenyo2012:amd-cluster}.  After commenting on the
experimental situation and interpretations for rotational bands in
$\isotope[10]{Be}$ (Sec.~\ref{sec:background}), we examine the \textit{ab
  initio} calculated rotational spectroscopy and the oscillator structure of the resulting
wave functions (Sec.~\ref{sec:spectrum}).

\section{Background: Experiment and cluster molecular interpretation}
\label{sec:background}

In $\isotope[10]{Be}$, the low-lying, positive-parity spectrum is known
experimentally to contain two $K^P=0^+$ bands: the ground state band and an
excited band at
$6.2\,\MeV$~\cite{freer2006:10be-resonant-molecule,bohlen2007:10be-pickup,suzuki2013:6he-alpha-10be-cluster}.
Both bands are observed up to their $4^+$ members.\footnote{\fnexptbandmembers}  The two bands have
significantly different moments of inertia, based on the observed band member
energies [$E(J)=E_0+AJ(J+1)$]:\footnote{\fnrotational} the excited band has a slope
($A\approx0.20\,\MeV$) lower than that of the ground state band
($A\approx0.59\,\MeV$) by about a factor of $3$, indicating approximately thrice the
moment of inertia. With its shallower slope, the excited
band becomes yrast at $J=4$.

Microscopic antisymmetrized molecular dynamics (AMD)
calculations~\cite{kanadaenyo1997:c-amd-pn-decoupling,kanadaenyo1999:10be-amd,suhara2010:amd-deformation}
support an interpretation in which $\isotope[10]{Be}$ may be described as a
molecule consisting of two $\alpha$ clusters (essentially an $\isotope[8]{Be}$
core) plus two ``valence'' neutrons which occupy molecular orbitals around these
clusters.  In the ground state band, these two neutrons occupy $\pi$ orbitals
(that is, with angular momentum projection $\pm1$ along the molecular axis),
loosely corresponding to a spherical shell model $0\hw$ configuration.  In the
excited band, these two neutrons occupy $\sigma$ orbitals (that is, with angular
momentum projection $0$ along the molecular axis), loosely corresponding to a
spherical shell model $2\hw$ configuration.  This configuration yields a larger
moment of inertia, due both to the greater spatial extent of the neutron
orbitals themselves along the molecular axis, and to an accompanying increase in
the inter-$\alpha$ separation~\cite{kanadaenyo1999:10be-amd}.

Furthermore, the AMD calculations suggest that the ground state of this nucleus
has triaxial
deformation~\cite{kanadaenyo1997:c-amd-pn-decoupling,suhara2010:amd-deformation},
specifically, proton-neutron triaxiality arising from the combination of an
overall prolate distribution for the protons and an overall oblate distribution
of neutrons, which are then oriented relative to each other such as to give an
overall triaxial shape.

In ideal triaxial rotation of an even-even nucleus, a more complex rotational
spectrum is expected~\cite{davydov1958:arm-intro,meyertervehn1975:triax-odda},
in which the $K=0$ ground state band is accompanied by $K=2,4,\ldots$ bands,
with $E2$ connections.  Indeed, the AMD calculations predict that a $K=2$ side
band should accompany the ground state band of
$\isotope[10]{Be}$~\cite{kanadaenyo1999:10be-amd}.  A possible (very short)
experimental side band has been proposed~\cite{bohlen2007:10be-pickup},
consisting of the experimental $2^+$ state at $5.96\,\MeV$ and a tentatively
identified $3^+$ member.  The resulting slope parameter ($A\approx0.57\,\MeV$)
is essentially identical to that of the ground state band.

\section{\textit{Ab initio} rotational spectrum}
\label{sec:spectrum}
\begin{figure}
\begin{center}
\includegraphics[width=\ifproofpre{0.80}{0.5}\hsize]{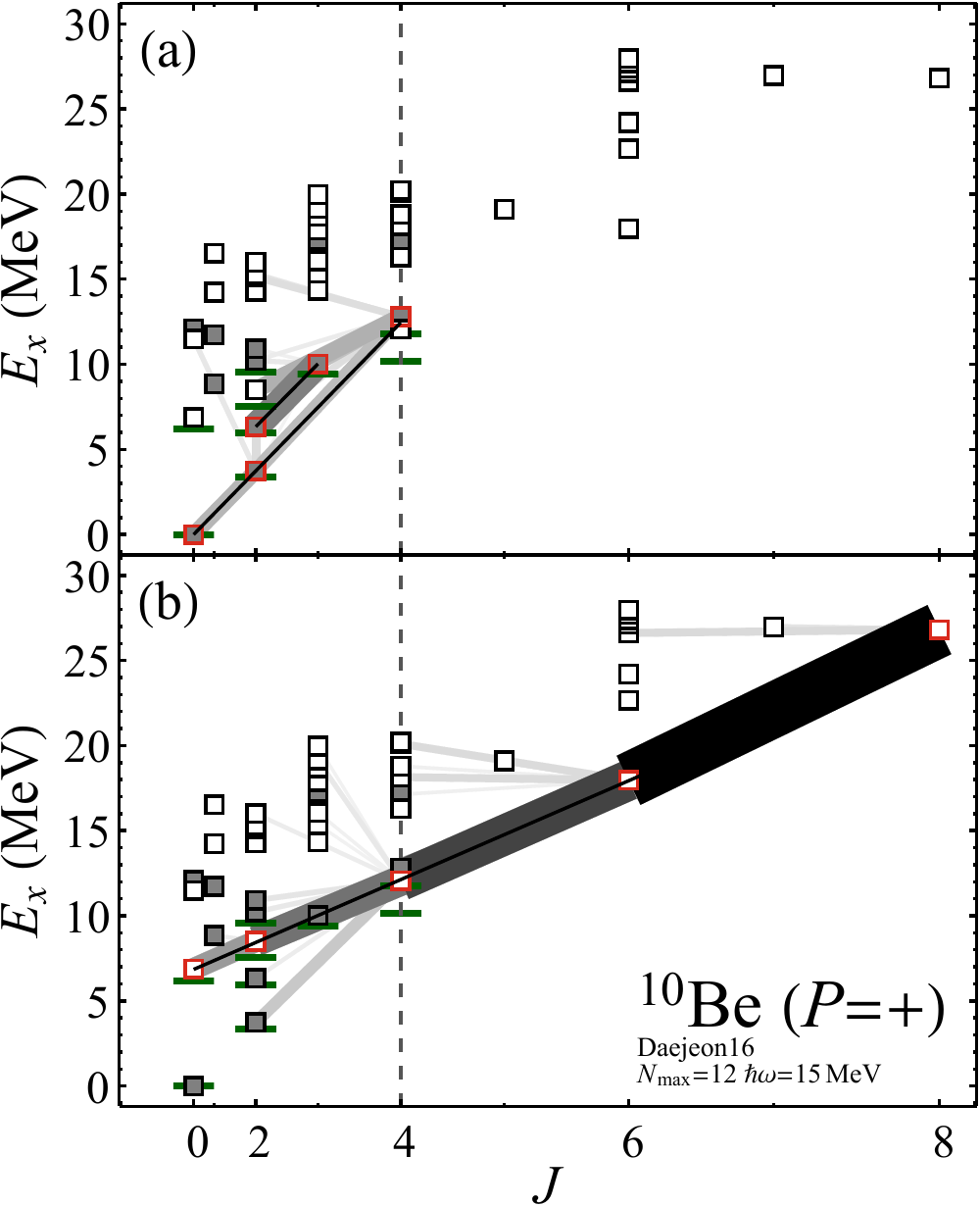}
\end{center}
\caption{\textit{Ab initio} calculated energy spectrum for $\isotope[10]{Be}$
  positive parity, showing $E2$ transitions originating from members of
  rotational bands: (a)~the ground state band ($K^P=0^+$) and side band
  ($K^P=2^+$) and (b)~the long band ($K^P=0^+$).  Rotational band members are
  highlighted (red squares), and rotational energy fits are indicated by lines.
  Experimental energies (green horizontal lines) are shown for comparison.
  States are approximately classified as $0\hw$ (filled symbols) or $2\hw$ (open
  symbols).  The $J$-decreasing $E2$ transitions originating from the rotational
  band members are shown (specifically, transitions with $J_f<J_i$ or with
  $J_f=J_i$ and $E_f<E_i$), with line thickness proportional to the $B(E2)$
  strength. The maximal angular momentum possible in the $0\hw$ valence space is indicated by the
  vertical dashed line. Calculation obtained for the Daejeon16 interaction, with oscillator
  basis parameter $\hw=15\,\MeV$ and truncation $\Nmax=12$.
\label{fig:levels-10be}
}
\end{figure}
\begin{figure}
\begin{center}
  \includegraphics[width=\ifproofpre{1.0}{0.5}\hsize]{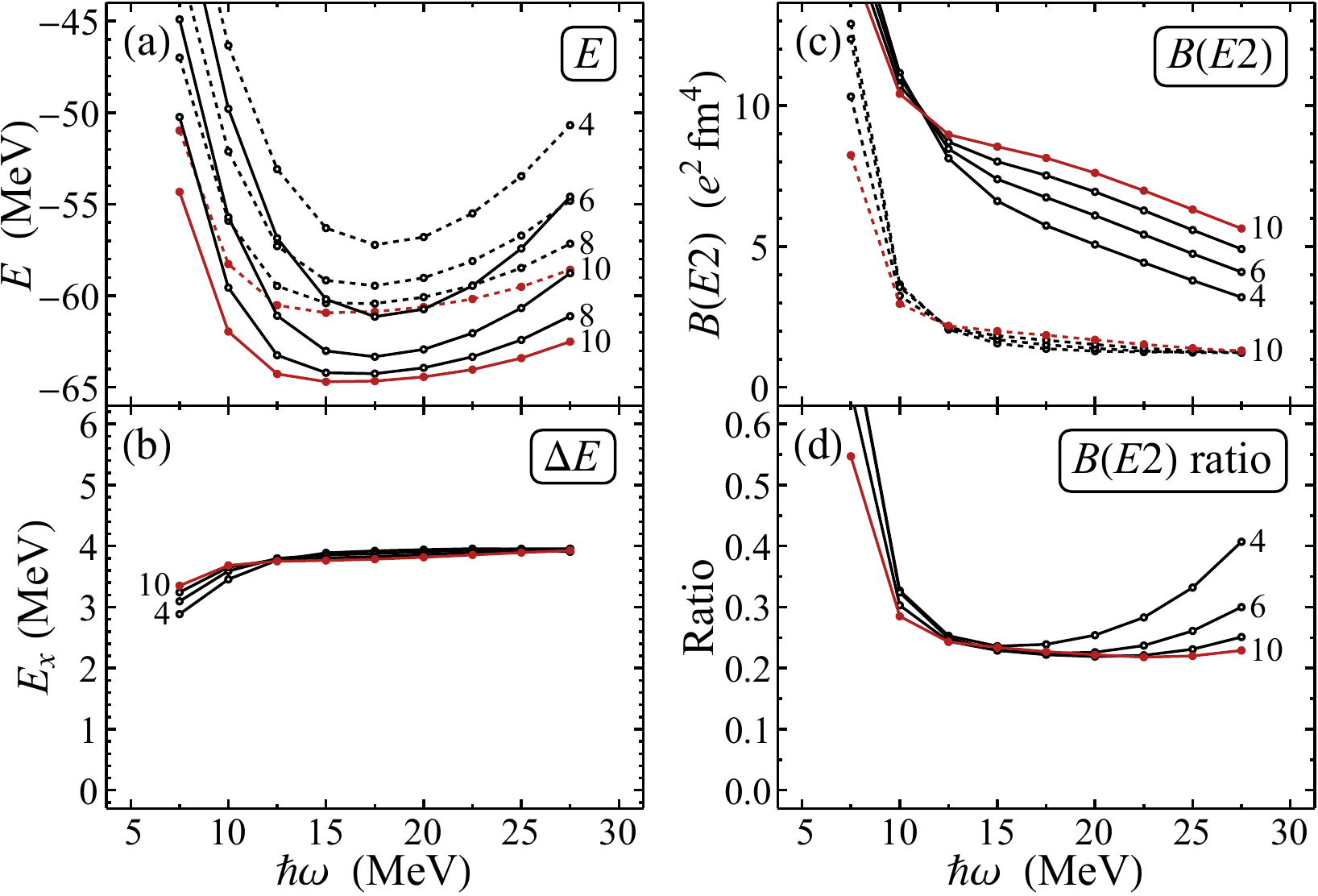}
\end{center}
\caption{Convergence of calculated energy and transition observables for
  $\isotope[10]{Be}$~(top) and corresponding relative observables~(bottom):
  (a)~energies of the $0^+_1$ ground state (solid curves) and $2^+_1$ rotational
  band member (dashed curves), (b)~the energy difference
  $E(2^+_1)-E(0^+_1)$, (c)~transition strengths $B(E2;2^+_1\rightarrow0^+_1)$
  (solid curves) and $B(E2;2^+_2\rightarrow2^+_1)$ (dashed curves) within the
  ground state rotational band and between the side band and the ground state
  band, respectively, (d)~the transition strength ratio
  $B(E2;2^+_2\rightarrow2^+_1)/B(E2;2^+_1\rightarrow0^+_1)$.
  Calculated values are shown
  as functions of the basis parameter $\hw$, for $\Nmax=4$ to $10$ (as labeled).
\label{fig:convergence-10be}
}
\end{figure}

In the NCCI approach, the nuclear many-body Hamiltonian is represented in terms
of a basis of antisymmetrized products (Slater determinants) of single-particle
states, typically harmonic oscillator states, and then diagonalized to yield the
energies and wave functions.  Any actual calculation must be carried out using a
finite, truncated basis, \textit{e.g.}, restricted to at most $\Nmax$ quanta of
excitation above the lowest Pauli-allowed filling of oscillator shells.

The excitation spectrum for the positive-parity states of $\isotope[10]{Be}$, as
obtained in an NCCI calculation with the Daejeon16 internucleon interaction,\footnote{\fndaejeon} is
shown in Fig.~\ref{fig:levels-10be}.  
The calculated levels (squares) are
overlaid with the experimental levels (horizontal
lines)~\cite{npa2004:008-010,freer2006:10be-resonant-molecule,bohlen2007:10be-pickup,suzuki2013:6he-alpha-10be-cluster}, and
energies are plotted against angular momentum scaled as $J(J+1)$ to facilitate
recognizing rotational bands.  These
calculations, obtained using the $M$-scheme
NCCI code
MFDn~\cite{maris2010:ncsm-mfdn-iccs10,aktulga2013:mfdn-scalability,shao2018:ncci-preconditioned},
are obtained using a harmonic oscillator basis with oscillator scale parameter
$\hw=15\,\MeV$ and truncation $\Nmax=12$.

The energies and wave functions obtained in any such truncated calculation
provide only an approximation to the true solution which would be obtained in
the full many-body space.  These results may be expected to converge towards the
full-space results with increasing $\Nmax$, but different aspects of the
calculated spectroscopy can have very different sensitivities, both to the
oscillator basis parameter $\hw$ and to $\Nmax$ itself, as illustrated in
Fig.~\ref{fig:convergence-10be}.

For instance, the calculated \textit{energy eigenvalues} themselves are changing
on a scale of $\MeV$ with each step in $\Nmax$
[Fig.~\ref{fig:convergence-10be}(a)], while many \textit{excitation energies}
(or relative energies, in general) are much better converged
[Fig.~\ref{fig:convergence-10be}(b)], on a scale of tens of $\keV$.  Similarly,
individual calculated $E2$ strengths do not typically approach a stable,
converged value in the computationally accessible spaces
[Fig.~\ref{fig:convergence-10be}(c)], yet many calculated relative transition
strengths, especially among members of the same rotational band or bands with
related structure, are much more stable [Fig.~\ref{fig:convergence-10be}(d)].
Thus, even when individual energies and transition strengths are not yet well
converged, rotational patterns involving relative energies and transition
strengths within bands are readily recognized (see discussions in
Refs.~\cite{caprio2015:berotor-ijmpe,caprio2019:bebands-ntse18,caprio20xx:bebands}).

Returning to the calculated positive-parity spectrum for $\isotope[10]{Be}$ in
Fig.~\ref{fig:levels-10be}, we may observe that the near-yrast states form three
rotational bands, in rough correspondence to the three experimental bands
(Sec.~\ref{sec:background}).  A $K^P=0^+$ ground state band ($0^+$, $2^+$,
$4^+$) is readily recognized in the calculations, from the rotational energies
[linear in $J(J+1)$] and enhanced $E2$ transitions
[Fig.~\ref{fig:levels-10be}(a)].  The moment of inertia ($A\approx0.62\,\MeV$)
is consistent with experiment.  The terminating angular momentum ($J=4$)
is also the maximal angular momentum which can be constructed for
$\isotope[10]{Be}$ in the $p$-shell valence space (or $0\hw$ space).

Then, the calculated yrare $2^+$ state and the first $3^+$ state are connected
by a strong $E2$ transition [Fig.~\ref{fig:levels-10be}(a)].  There are also
significant $E2$ strengths from these states to the ground state band.  These
states thus form a putative short $K^P=2^+$ band, a ``side band''
(spectroscopically speaking) to the ground state band. The calculated band head is at just over $6\,\MeV$
excitation energy, while the energy difference gives a moment of inertia
($A\approx0.62\,\MeV$) essentially identical to that of the ground state band.

The enhanced transitions between the side band and ground state band are,
specifically, the $2^+_{K=2}\rightarrow 2^+_{K=0}$ and $3^+_{K=2}\rightarrow
4^+_{K=0}$ transition.  This transition pattern is not at all what would be
expected from the Alaga rules~\cite{alaga1955:branching} for transitions between
true $K=2$ and $K=0$ bands, as obtained for an axially symmetric rotor.  For instance, the
$2_{K=2}\rightarrow2_{K=0}$ and $2_{K=2}\rightarrow0_{K=0}$ transitions should
be of comparable strength
[$B(E2;2_{K=2}\rightarrow2_{K=0})/B(E2;2_{K=2}\rightarrow0_{K=0})\approx1.4$]
for an axially symmetric rotor, while the computed $2^+_{K=2}\rightarrow
0^+_{K=0}$ strength here is negligible.  Rather, the calculated transitions
follow the $\gamma$-parity~\cite{bes1959:gamma} selection rules expected for a
$\gamma=30^\circ$ triaxial rotor (see, \textit{e.g.}, Fig.~17 of
Ref.~\cite{caprio2005:ibmpn2}), where only states of opposite $\gamma$ parity
are connected by $E2$ transitions.

Finally, an excited $K^P=0^+$ band starts from the calculated $0^+_2$ state as
its band head [Fig.~\ref{fig:levels-10be}(b)] and extends through an $8^+$
member.  The excited band members have essentially negligible $E2$ connections
to the ground state band members.  (An exception arises since the $4^+$ members
of the ground and excited $K^P=0^+$ bands undergo transient two-state mixing,
when their calculated energies cross at $\Nmax=10$ and $12$, as discussed below.
This mixing gives rise to modest transitions connecting the $4^+$ and $2^+$
members of the different bands, seen in Fig.~\ref{fig:levels-10be}.)

The excited $K^P=0^+$ band has a much shallower slope ($A\approx0.26\,\MeV$), and thus
larger moment of inertia, than the ground state band, as in experiment.  This
larger deformation is already suggested by the greater in-band $E2$ strengths
[comparing Fig.~\ref{fig:levels-10be}(a) and Fig.~\ref{fig:levels-10be}(b)].
However, $E2$ observables are only partially sensitive to this deformation, as
it appears to largely arise from the neutrons.  Comparing the calculated
quadrupole moment of the proton distribution (\textit{i.e.}, the physical
electric quadrupole moment) with that of the neutron distribution gives a ratio
of merely $Q_n/Q_p\approx 0.7$ for the ground state band, but a much larger
ratio $Q_n/Q_p\approx 2.1$ for the excited $K^P=0^+$ band (see also Fig.~18 of
Ref.~\cite{maris2015:berotor2}).

\begin{figure}
\begin{center}
\includegraphics[width=\ifproofpre{0.80}{1}\hsize]{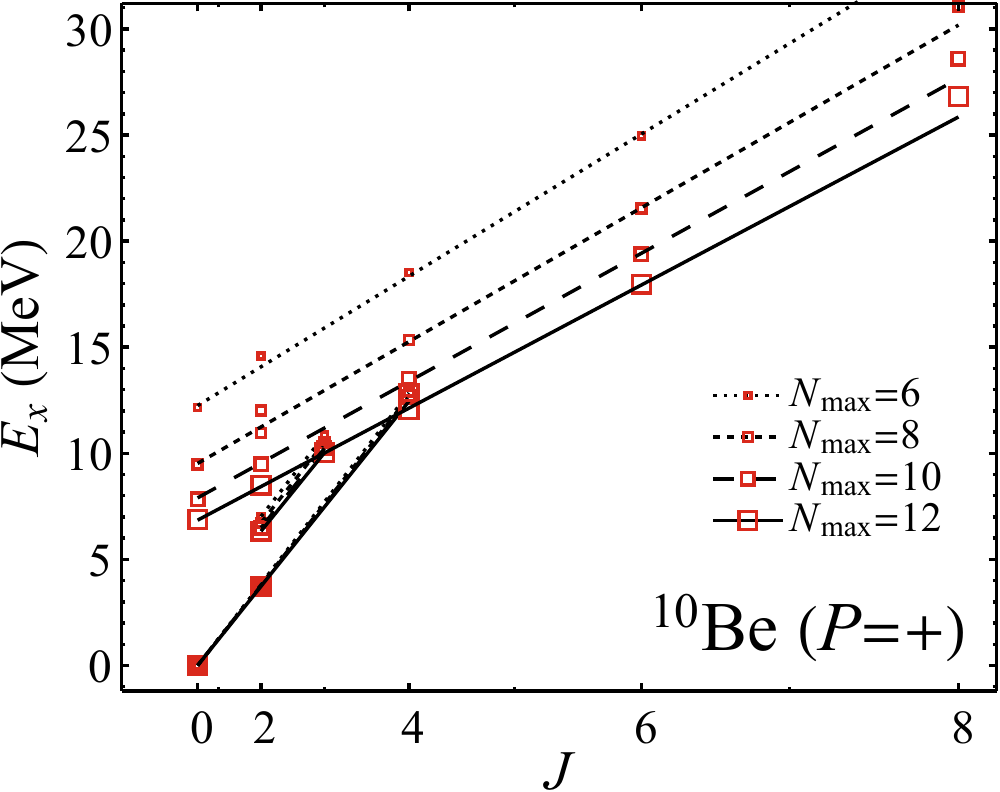}
\end{center}
\caption{Calculated excitation energies for rotational band members in
  $\isotope[10]{Be}$, for $\Nmax=6$ to $12$ (dotted through solid curves).
\label{fig:band-energies-10be}
}
\end{figure}

Before attempting any meaningful comparison with experiment, we must take into
account the convergence of the calculated energies of the band members, as
explored for $\Nmax=6$ to $12$ in Fig.~\ref{fig:band-energies-10be}.
Within the ground state band, the excitation
energies have minimal $\Nmax$ dependence, although the slope does slightly
decrease (from $A\approx0.64\,\MeV$ to $0.62\,\MeV$) over this range of $\Nmax$.
The side band descends in excitation energy by $\lesssim1\,\MeV$, indeed, moving closer to experiment
[Fig.~\ref{fig:levels-10be}(a)].
In contrast, the calculated excitation energies of the excited $K^P=0^+$ band
members plummet by more than $5\,\MeV$.  The ground
state and excited band $4^+$ members cross, so that the excited band $4^+$
member becomes yrast, as in experiment, at $\Nmax=12$.  The calculated excited band members at
$\Nmax=12$ still lie above experiment, by
$\sim0.7\,\MeV$ for the band head and $\sim2\,\MeV$ for the $4^+$ member [Fig.~\ref{fig:levels-10be}(b)].

\begin{figure}
\begin{center}
\includegraphics[width=\ifproofpre{0.75}{1}\hsize]{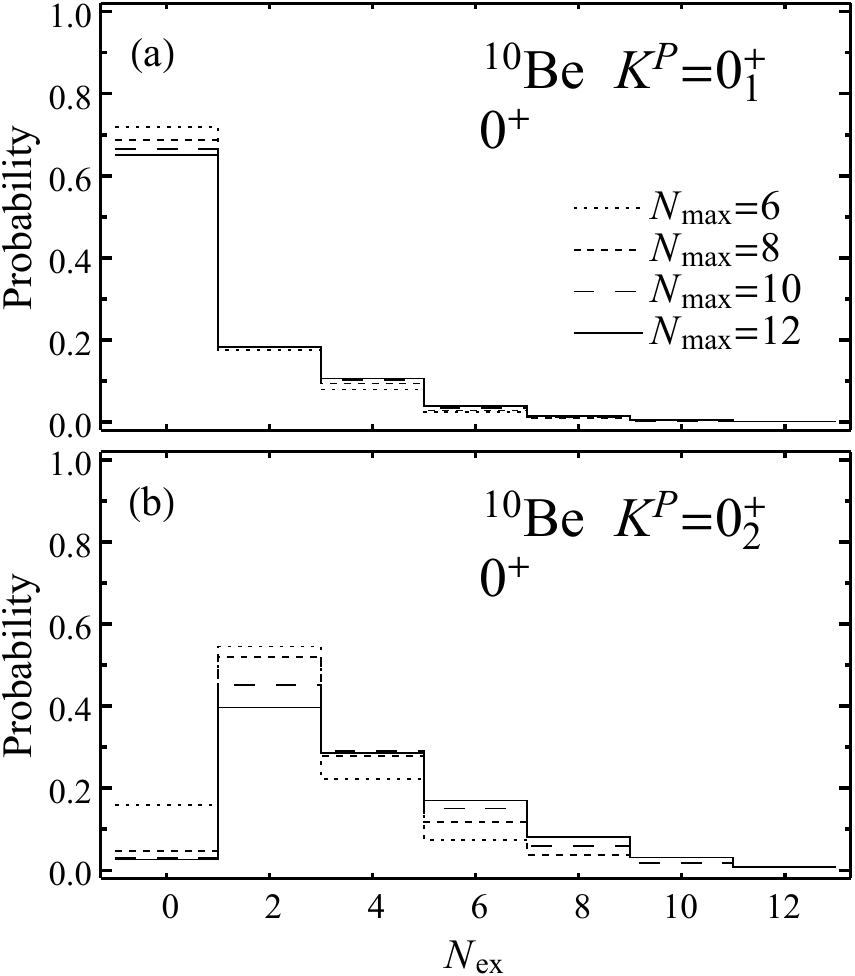}
\end{center}
\caption{Decompositions of $\isotope[10]{Be}$ rotational band heads in
  oscillator space: (a)~the $0^+$ ground state and (b)~the $0^+$ band head of
  the long band.  Shown for $\Nmax=6$ to $12$ (dotted through solid curves).
\label{fig:decompositions-Nex-10be}
}
\end{figure}

In a conventional shell-model description, based on interacting nucleons in a
harmonic-oscillator mean-field potential, there is a natural organization of
states into $0\hw$ and $2\hw$ states based on the predominant number of
oscillator excitations above the valence shell.  Such a classification seems to
also be relevant to nuclear states obtained in \textit{ab initio} approaches,
despite there being no explicitly imposed mean field.

In particular, the decomposition of the calculated wave function into the
contributions coming from oscillator basis configurations with different numbers of
oscillator quanta (\textit{e.g.}, Ref.~\cite{caurier2002:ncsm-a10}) can be
suggestive of a $0\hw$ or $2\hw$ nature, though this interpretation is at best
approximate (see Ref.~\cite{caprio20xx:bebands} for discussion).  To explore the
oscillator structure of the $\isotope[10]{Be}$ rotational bands, let us examine such
decompositions, that is, by the number $\Nex$ of excitation quanta relative to the lowest
Pauli allowed oscillator configuration, as shown in
Fig.~\ref{fig:decompositions-Nex-10be} for the $K^P=0^+$ band head states.

The predominant contribution
to the ground state comes from $\Nex=0$ (or $0\hw$) configurations
[Fig.~\ref{fig:decompositions-Nex-10be}(a)].  Similar decompositions are found
for the other members of the ground state band and side band, suggesting that
their essential structure can largely be described in a $0\hw$ shell model.

In contrast, the excited $K^P=0^+$ band could not possibly be explained entirely
within the $0\hw$ shell model, as we may recall it extends beyond the maximal
valence angular momentum ($J=4$).  For the excited $0^+$ band head
[Fig.~\ref{fig:decompositions-Nex-10be}(b)], the predominant contributions come
from $\Nex=2$ ($2\hw$) and higher configurations, falling off gradually with
higher $\Nex$, while the $\Nex=0$ contribution is just a few percent.  Similar
decompositions are found for the other excited $K^P=0^+$ band members.

\section{Conclusion}
\label{sec:concl}

\textit{Ab initio} NCCI calculations of $\isotope[10]{Be}$ were already known to
be suggestive of rotational
structure~\cite{caprio2013:berotor,maris2015:berotor2}.  However, the present
calculations with the Daejeon16 interaction provide a picture which is
qualitatively and, to within the limitations of convergence, quantitatively
consistent with experiment and which would seem to be qualitatively consistent
with the expectations from microscopic cluster molecular descriptions as well.

Two coexisting rotational structures are found at low energy in the
$\isotope[10]{Be}$ positive parity space.  The ground state band ($K^P=0^+$) is
predicted to be accompanied by a side band ($K^P=2^+$) of similar moment of
inertia, both having essentially $0\hw$ structure.  The $E2$ transition pattern
is more consistent with triaxial structure than with axially symmetric rotation.
Then, a low-lying $K^P=0^+$ excited band, with much larger moment of inertia,
although only experimentally observed through $J=4$, is expected to extend to
higher angular momentum.  This band, in contrast, has something akin to $2\hw$
structure, and its greater deformation arises primarily from the neutrons.


\section*{Acknowledgements}

This material is based upon work supported by the U.S.~Department
of Energy, Office of Science, under Award Numbers DE-FG02-95ER-40934,
DESC00018223 (SciDAC4/NUCLEI), and DE-FG02-87ER40371, and by the U.S.~National
Science Foundation under Award Number NSF-PHY05-52843.  TRIUMF receives federal
funding via a contribution agreement with the National Research Council of
Canada.  This research used computational resources of the University of Notre
Dame Center for Research Computing and of the National Energy Research
Scientific Computing Center (NERSC), a U.S.~Department of Energy, Office of
Science, user facility supported under Contract~DE-AC02-05CH11231.

%


\end{document}